\documentclass[aps,prb,twocolumn,amssymb,showpacs,amsmath,superscriptaddress,floatfix]{revtex4}
\usepackage{graphicx}
\usepackage{bm}
\usepackage{multirow}
\usepackage{appendix}

\begin{document}
\title {Impact of the orbital current order on the superconducting properties of the kagome superconductors}

\author{Hong-Min Jiang}
\email{monsoonjhm@sina.com} \affiliation{School of Science, Zhejiang
University of Science and Technology, Hangzhou 310023, China}
\author{Ming-Xun Liu}
\affiliation{School of Science, Zhejiang University of Science and
Technology, Hangzhou 310023, China}
\author{Shun-Li Yu}
\email{slyu@nju.edu.cn} \affiliation{School of Physics, National
Laboratory of Solid State Microstructures, Nanjing University,
Nanjing 210093, China}

\date{\today}

\begin{abstract}
Motivated by recent experimental evidences signalling the chiral
charge order in the vanadium-based kagome superconductors, we
theoretically investigate the impact of the chiral flux charge order
over the experimental outcomes for the normal and the SC properties.
It is revealed that the spectral weight on the Fermi surface (FS) is
partially gaped by the chiral flux charge order with the reservation
of the spectral weight on the $M$ points and the midpoint between
the two adjacent $M$ points, resulting in the momentum-dependent
energy gap being consistent with the recent experimental
observations. More importantly, by considering the influence of the
chiral flux charge order, we find that a conventional fully gapped
SC pairing state evolves into a nodal gap feature for the spectral
weight due to the spectral gap modulations on the FS. As a result,
the U-shaped density of states (DOS) deforms to the V-shaped one
along with the residual DOS near the Fermi energy. These results
bear some resemblance to the experimental observations, and may
serve as a promising proposal to mediate the divergent or seemingly
contradictory experimental outcomes about the SC pairing symmetry.
\end{abstract}

\maketitle

\section{introduction}

The recent discovery of superconductivity in a family of compounds
AV$_{3}$Sb$_{5}$ (A=K, Rb, Cs), which share a common lattice
structure with kagome net of vanadium atoms, has triggered a new
surge of interest in the investigation of
superconductivity~\cite{Ortiz1,SYYang1,Ortiz2,QYin1,KYChen1,YWang1,ZZhang1,YXJiang1,
FHYu1,XChen1,HZhao1,HChen1,HSXu1,Liang1,CMu1,CCZhao1,SNi1,WDuan1,
Xiang1,PhysRevX.11.041030,PhysRevB.104.L041101,
PhysRevX.11.041010,NatPhys.M.Kang,YFu1,HTan1,Shumiya1,FHYu2,LYin1,Nakayama2,
KJiang1,LNie1,HLuo1,Neupert1,YSong1,Nakayama1,HLi1,CGuo1}. The
fascinating aspects of these materials lie in the exotic quantum
physics as they integrate with the geometrical lattice frustration,
the van Hove filling, the nontrivial band topology, and the
interplay between the charge density wave and
superconductivity~\cite{HChen1,SLYu1,Kies1,WSWang1,Kies2}, which
makes the emergence of superconductivity in these materials is in
themselves exotic and rare.

A central issue about the superconductivity is to unveil the
superconducting (SC) pairing mechanism. To search for clues to this
puzzling question, the determination of the pairing symmetry of the
SC order parameter is thought to be a prerequisite step. However,
the inconsistent or even contradicting results have been found so
far in experimental measurements and data analyses. The temperature
dependence of the nuclear spin-lattice relaxation rate shows a
Hebel-Slichter coherence peak just below $T_{c}$, indicating that
CsV$_{3}$Sb$_{5}$ is a conventional $s$-wave
superconductor~\cite{CMu1}. The penetration depth measurements also
lend support to such a view~\cite{WDuan1}. Nevertheless, the
measurements of thermal conductivity on CsV$_{3}$Sb$_{5}$ at
ultra-low temperature evidenced a finite residual linear term,
pointing to an unconventional nodal SC gap~\cite{CCZhao1}. In
accordance with this, the V-shaped SC gaps with residual zero-energy
density of states (DOS) also suggest an anisotropic SC gap with
nodes~\cite{HSXu1,Liang1,HChen1}. Moreover, the scanning tunnelling
microscope/spectroscopy (STM/STS) experiment on CsV$_{3}$Sb$_{5}$ at
ultra-low temperature revealed a two-gap structure with multiple
sets of coherent peaks and residual zero-energy DOS, accompanied by
the magnetic/non-magnetic impurity effects, implying a rather novel
and interesting SC gap, i.e., the sign preserved multiband
superconductivity with gap nodes~\cite{HSXu1}.

Theoretically, the results about the pairing symmetries vary widely
as well, depending on the different methods and models. Early before
the discovery of the superconductivity in AV$_{3}$Sb$_{5}$, while a
chiral $d_{x^{2}-y^{2}}+id_{xy}$-wave SC state was predicted to be
the most favorable one within a reasonable parameter range for the
van Hove filling kagome system based on the variational cluster
approach and the perturbative renormalization group analysis to a
single-orbital Hubbard model~\cite{SLYu1,Kies1}, the singular-mode
functional renormalization group theory discovered a rich variety of
electronic instabilities ranging form $s$-wave and $d$-wave to
$d_{x^{2}-y^{2}}+id_{xy}$-wave superconductivities under short-range
interactions~\cite{WSWang1}. Later on, with the inspiration from the
discovery of superconductivity in the family of AV$_{3}$Sb$_{5}$, a
random phase approximation based on a two-orbital model revealed an
$f$-wave pairing instability over a large range of coupling
strength, succeeded by $d$-wave singlet pairing for stronger
coupling~\cite{XWu1}. On the other hand, the determinant quantum
Monte Carlo calculations on the kagome-lattice Hubbard model found
the dominating pairing channel was $d_{x^{2}-y^{2}}+id_{xy}$
($s_{ex}$)-wave in the hole (electron)-doped case~\cite{CWen1}. What
is more, it has been uncovered that the mechanism of bond-order
fluctuations could give rise to both singlet $s$-wave and triplet
$p$-wave superconductivity~\cite{Tazai1}. Nevertheless, the
experimental controversy concerning the SC pairing symmetry in the
family of AV$_{3}$Sb$_{5}$ remains unsettled.

One of the reasons for the divergent experimental outcomes lies in
that the SC order may intertwine with other unconventional electron
orders, adding another layer of complexity to this already
challenging issue in the field. Although the possibility of
long-range magnetic order in the AV$_{3}$Sb$_{5}$ crystal has been
ruled out by the neutron scattering~\cite{Ortiz3} and muon spin
spectroscopy~\cite{Kenney1} measurements, a giant anomalous Hall
effect has still been observed above the entrance of the SC state
with the concomitant onset of a $2\times 2$ charge density wave
(CDW) order~\cite{SYYang1,FHYu1}, indicating this time-reversal
symmetry-breaking transition derives primarily from the charge
degree of freedom~\cite{Kenney1}. So far, there are an increasing
number of experimental evidences supporting that the CDW state has a
$2\times2$ chiral flux
order~\cite{YXJiang1,Shumiya1,CGuo1,Mielke1,LYu1,YXu1,XZhou1,DChen1,QWu1,YHu1}.
Furthermore, the muon spin relaxation technic observed a noticeable
enhancement of the internal field width, which takes place just
below the charge ordering temperature and persists into the SC
state~\cite{Mielke1}, pointing to time-reversal symmetry-breaking
charge order intertwining with unconventional superconductivity.
Thus, the chiral CDW should be considered in analyzing the SC
properties in the vanadium-based kagome superconductors.

Given the fairly good Fermi surface (FS) nesting and proximity to
the von Hove singularity, the system is prone to the instability of
``triple-$Q$" CDW at the three nesting wave
vectors~\cite{Vender1,YPLin1,TPark1,YPLin2,Feng1,Feng2,Denner1,YGu1,YPLin3}.
While the real component of the ``triple-$Q$" bond charge order
conforms to the modulated superlattice pattern uncovered in the
experiments, the theoretical proposal of ``triple-$Q$" imaginary
CDWs, which was originally put forward on the honeycomb and
triangular lattices~\cite{Vender1,YPLin1}, appears consistent
simultaneously with the superlattice modulations and the
time-reversal symmetry breaking. Subsequently, various chiral flux
phases with different configurations of orbital current have been
proposed in a series of theoretical studies on the kagome
superconductors~\cite{TPark1,YPLin2,Feng1,Feng2,Denner1,YGu1}. A
recent self-consistent mean-field study has shown that the chiral
CDW in Fig. 1(a) can be stabilized by moderately small intersite
Coulomb interactions and it might be the most relevant time-reversal
symmetry-breaking state in AV$_{3}$Sb$_{5}$~\cite{JWDong1}. In
addition, the coexistence of the chiral CDW with a conventional
fully gapped superconductivity could lead to gapless edge modes on
the domains of the lattice symmetry breaking order~\cite{YGu1}.
These edge modes with gapless excitations could account for the
residual DOS and the finite residual thermal conductivity. However,
there is still lack of a systematic investigation on the direct
influence of the chiral CDW over the SC pairing properties,
especially when one considers the growing experimental evidences
pointing to the persistence of the time-reversal symmetry-breaking
CDW well into the SC state~\cite{YXJiang1,CGuo1,LZheng1}.

In this paper, we aim to fill up the blank by showing that the
chiral flux phase in Fig.~\ref{fig1}(a) as a representative of the
time-reversal symmetry-breaking $2\times 2$ CDW has a profound
impact on the experimental outcomes with respect to both the normal
and the SC properties. By unfolding the spectral weight function and
the energy bands to the primitive Brillouin zone (PBZ), we show that
only portions of the FS are gaped by the CDW order. While the
original band near the saddle point is split into three sub-bands by
the CDW order, three new saddle points emerge in the CDW phase, and
especially there is still an obvious residual saddle-point spectrum
at the Fermi level. Importantly, the novelty we discovered in the
calculations is such that a conventional fully gapped SC pairing
state will acquire a nodal gap feature for the spectral weight on
the FS when the impact of the chiral flux CDW order is taken into
account. The nodal gap feature manifests itself as the evolution
from the U-shaped DOS to the V-shaped one along with the residual
DOS near the Fermi energy, which is the direct outcome of the charge
order induced gap modulations of the spectral function on the FS.
These results not only account for some experimental observations,
but also provide an alternative scenario to reconcile the divergent
or seemingly contradictory experimental outcomes regarding the SC
pairing symmetry.

The remainder of the paper is organized as follows. In Sec. II, we
introduce the model Hamiltonian and carry out analytical
calculations. In Sec. III, we present numerical calculations and
discuss the results. In Sec. IV, we make a conclusion.

\section{model and method}
The $2\times 2$ CDW order quadruply enlarges the unit cell, as
indicated by the dashed lines in Fig.~\ref{fig1}(a). Among the
possible time-reversal symmetry-breaking charge order
configurations, the chiral flux phase with configuration of the
orbital current shown in Fig.~\ref{fig1}(a) has been found to be
energetically favorable~\cite{Feng2}. Thus, we adopt this typical
CDW configuration as a representative of the demonstration, and
leave other possibilities for future researches. It is widely
believed that the inter-scattering between three van Hove points
with wave vectors $\mathbf{Q}_{a}=(0,2\pi/\sqrt{3})$,
$\mathbf{Q}_{b}=(-\pi,-\pi/\sqrt{3})$ and
$\mathbf{Q}_{c}=(\pi,-\pi/\sqrt{3})$ causes the CDW in
AV$_{3}$Sb$_{5}$ [The wave vectors are shown in
Fig.~\ref{fig1}(b).]. Meanwhile, the van Hove filling was also
proposed to be crucial to the superconductivity in AV$_{3}$Sb$_{5}$.
A single orbital tight binding model near the van Hove filling
produces the essential feature of the FS and the van Hove physics.
Therefore, to capture the main physics of the topological CDW and
its impacts on the SC in AV$_{3}$Sb$_{5}$, we adopt a minimum single
orbital model.

The single orbital model involving the effective electron hoppings
on a kagome lattice can be described by the following tight-binding
Hamiltonian,
\begin{eqnarray}
H_{0}&=&-t\sum_{\langle
\textbf{ij}\rangle\sigma}(c^{\dag}_{\textbf{i}\sigma}c_{\textbf{j}\sigma}+h.c.)
-\mu\sum_{\textbf{i}\sigma}c^{\dag}_{\textbf{i}\sigma}c_{\textbf{i}\sigma},
\end{eqnarray}
where $c^{\dag}_{\textbf{i}\sigma}$ creates an electron with spin
$\sigma$ on the site $\mathbf{r}_{i}$ of the kagome lattice and
$\langle\textbf{ij}\rangle$ denotes nearest-neighbors (NN). $t$ is
the hopping integral between the NN sites, and $\mu$ stands for the
chemical potential. The Hamiltonian $H_{0}$ can be written in the
momentum space as,
\begin{eqnarray}
H_{0}(\mathbf{k})&=&\sum_{\mathbf{k}\sigma}\hat{\Psi}^{\dag}_{\mathbf{k}\sigma}\hat{\mathcal{H}}^{0}_{\mathbf{k}}
\hat{\Psi}_{\mathbf{k}\sigma},
\end{eqnarray}
with
$\hat{\Psi}_{\mathbf{k}\sigma}=(c_{A\mathbf{k}\sigma},c_{B\mathbf{k}\sigma},c_{C\mathbf{k}\sigma})^{T}$
and
\begin{eqnarray}
\hat{\mathcal{H}}^{0}_{\mathbf{k}}=\left(
\begin{array}{ccc}
-\mu & -2t\cos k_{1} &
-2t\cos k_{2} \\
-2t\cos k_{1} & -\mu & -2t\cos k_{3} \\
-2t\cos k_{2} & -2t\cos k_{3} & -\mu
\end{array}
\right).
\end{eqnarray}
The index $m=A,B,C$ in $c_{mk\sigma}$ labels the three basis sites
in the triangular primitive unit cell (PUC). $k_{n}$ is abbreviated
from $\mathbf{k}\cdot\mathbf{\tau}_{n}$ with
$\mathbf{\tau}_{1}=\hat{x}/2$,
$\mathbf{\tau}_{2}=(\hat{x}+\sqrt{3}\hat{y})/4$ and
$\mathbf{\tau}_{3}=\mathbf{\tau}_{2}-\mathbf{\tau}_{1}$ denoting the
three NN vectors. The spectral function of $H_{0}(\mathbf{k})$
defined as
$A^{0}(\mathbf{k},E)=-\frac{1}{\pi}\textmd{Tr}[\textmd{Im}\hat{G}^{0}(\mathbf{k},iE\rightarrow
E+i0^{+})]$ with
$\hat{G}^{0}(\mathbf{k},iE)=[iE\hat{I}-\hat{\mathcal{H}}^{0}_{\mathbf{k}}]^{-1}$.
Near the van Hove filling with $1/6$ hole doping, the spectral
function at zero energy $E=0$ produces the hexagonal FS and the van
Hove singularities at $M$ points, as shown in Fig.~\ref{fig1}(b),
which produce the essences of the FS and energy band observed in the
angle-resolved photoemission spectroscopy (ARPES) experiment and the
density functional theory calculations~\cite{Ortiz1}.

The second part of the Hamiltonian accounts for the chiral flux CDW
order,
\begin{eqnarray}
H_{C}=&&i\lambda\sum_{\langle\textbf{ij}\rangle
\sigma}\eta_{\mathbf{ij}}(c^{\dag}_{\mathbf{i}\sigma}c_{\mathbf{j}\sigma}-h.c.),
\end{eqnarray}
where $\lambda$ denotes the strength of the orbital current order,
and $\eta_{\mathbf{ij}}=+1$ if the hopping is positioned in the same
direction of the orbital current and otherwise
$\eta_{\mathbf{ij}}=-1$.

The third term accounts for the SC pairing. It reads
\begin{eqnarray}
H_{P}&=&\sum_{\mathbf{i}}(\Delta
c^{\dag}_{\mathbf{i}\uparrow}c^{\dag}_{\mathbf{i}\downarrow} +h.c.).
\end{eqnarray}
The on-site $s$-wave SC order parameter $\Delta=-V\langle
c_{\textbf{i}\uparrow}c_{\textbf{i}\downarrow}\rangle$ is assumed to
derive from the effective interaction between electrons. In the
calculations, we choose the typical values of the effective pairing
interaction $V=1.6$. Varying the pairing interaction will change the
pairing amplitude, but the results presented here will be
qualitatively unchanged if the CDW order strength changes in
parallel.

In the coexistence of SC and chiral flux $2\times 2$ CDW orders, the
total Hamiltonian $H=H_{0}+H_{P}+H_{C}$ can be written in the
momentum space within one enlarged unit cell (EUC) shown in
Fig.~\ref{fig1}(a) as,
\begin{eqnarray}
H(\mathbf{k})=&&-t\sum_{\mathbf{k},\langle\mathbf{\tilde{i}\tilde{j}}\rangle,\sigma}[c^{\dag}_{\mathbf{k}\mathbf{\tilde{i}}\sigma}c_{\mathbf{k}\mathbf{\tilde{j}}\sigma}e^{-i\mathbf{k}\cdot(\mathbf{r}_{\tilde{i}}-\mathbf{r}_{\tilde{j}})}+h.c.]
\nonumber\\
&&-\mu\sum_{\mathbf{k},\mathbf{\tilde{i}},\sigma}c^{\dag}_{\mathbf{k}\mathbf{\tilde{i}}\sigma}c_{\mathbf{k}\mathbf{\tilde{i}}\sigma} \nonumber\\
&&+i\lambda\sum_{\mathbf{k},\langle\mathbf{\tilde{i}\tilde{j}}\rangle,\sigma}\eta_{\mathbf{\tilde{i}\tilde{j}}}[c^{\dag}_{\mathbf{k}\mathbf{\tilde{i}}\sigma}c_{\mathbf{k}\mathbf{\tilde{j}}\sigma}e^{-i\mathbf{k}\cdot(\mathbf{r}_{\tilde{i}}-\mathbf{r}_{\tilde{j}})}-h.c.]
\nonumber\\
&&+\sum_{\mathbf{k},\mathbf{\tilde{i}}}(\Delta
c^{\dag}_{\mathbf{k}\mathbf{\tilde{i}}\uparrow}c^{\dag}_{-\mathbf{k}\mathbf{\tilde{i}}\downarrow}
+h.c.),
\end{eqnarray}
where $\mathbf{\tilde{i}}\in \textmd{EUC}$ represents the lattice
site being within one EUC, and
$\langle\mathbf{\tilde{i}\tilde{j}}\rangle$ denotes the NN sites
with the periodic boundary condition implicitly assumed.
Accordingly, the summation of $\mathbf{k}$ should be in principle
over the reduced Brillouin zone (RBZ) as enveloped by the white
dashed lines in Fig.~\ref{fig1}(c).

Based on the Bogoliubov transformation, we obtain the following
Bogoliubov-de Gennes equations in the EUC,
\begin{eqnarray}
\sum_{\mathbf{k}}\sum_{\mathbf{\tilde{j}}}\left(
\begin{array}{lr}
H_{\mathbf{\tilde{i}\tilde{j}},\sigma} &
\Delta_{\mathbf{\tilde{i}\tilde{j}}} \\
\Delta^{\ast}_{\mathbf{\tilde{i}\tilde{j}}} &
-H^{\ast}_{\mathbf{\tilde{i}\tilde{j}},\bar{\sigma}}
\end{array}
\right)\exp[i\mathbf{k}\cdot(\textbf{r}_{\mathbf{\tilde{j}}}-\textbf{r}_{\mathbf{\tilde{i}}})]\left(
\begin{array}{lr}
u^{\mathbf{k}}_{n,\mathbf{\tilde{j}},\sigma} \\
v^{\mathbf{k}}_{n,\mathbf{\tilde{j}},\bar{\sigma}}
\end{array}
\right) \nonumber\\
= E^{\mathbf{k}}_{n}\left(
\begin{array}{lr}
u^{\mathbf{k}}_{n,\mathbf{\tilde{i}},\sigma} \\
v^{\mathbf{k}}_{n,\mathbf{\tilde{i}},\bar{\sigma}}
\end{array}
\right),
\end{eqnarray}
where
$H_{\mathbf{\tilde{i}}\mathbf{\tilde{j}},\sigma}=(-t+i\lambda\eta_{\mathbf{\tilde{i}\tilde{j}}})\delta_{\mathbf{\tilde{i}}+\mathbf{\tau}_{\mathbf{\tilde{j}}},\mathbf{\tilde{j}}}-\mu\delta_{\mathbf{\tilde{i}},\mathbf{\tilde{j}}}$
with $\mathbf{\tau}_{\mathbf{\tilde{j}}}$ denoting the four NN
vectors and
$\Delta_{\mathbf{\tilde{i}\tilde{j}}}=\Delta\delta_{\mathbf{\tilde{i}},\mathbf{\tilde{j}}}$.
$u^{\mathbf{k}}_{n,\mathbf{\tilde{i}},\sigma}$ and
$v^{\mathbf{k}}_{n,\mathbf{\tilde{i}},\bar{\sigma}}$ are the
Bogoliubov quasiparticle amplitudes on the $\mathbf{\tilde{i}}$-th
site with corresponding momentum $\mathbf{k}$ and eigenvalue
$E^{\mathbf{k}}_{n}$. The SC pairing amplitude and electron
densities are obtained through the following self-consistent
equations,
\begin{eqnarray}
\Delta=&&\frac{V}{2}\sum_{\mathbf{k},n}u^{\mathbf{k}}_{n,\mathbf{\tilde{i}},\sigma}
v^{\mathbf{k}\ast}_{n,\mathbf{\tilde{i}},\bar{\sigma}}
\tanh(\frac{E^{\mathbf{k}}_{n}}{2k_{B}T}) \nonumber\\
n_{\mathbf{\tilde{i}}}=&&\sum_{\mathbf{k},n}\{|u^{\mathbf{k}}_{n,\mathbf{\tilde{i}},\uparrow}|^{2}f(E^{\mathbf{k}}_{n})+|v^{\mathbf{k}}_{n,\mathbf{\tilde{i}},\downarrow}|^{2}[1-f(E^{\mathbf{k}}_{n})]\}.
\end{eqnarray}

Then, the single particle Green functions
$G_{\mathbf{\tilde{i}}\mathbf{\tilde{j}}}(\mathbf{k},i\omega)=-\int^{\beta}_{0}d\tau\exp^{i\omega\tau}\langle
T_{\tau}c_{\mathbf{k}\mathbf{\tilde{i}}}(i\tau)c^{\dag}_{\mathbf{k}\mathbf{\tilde{j}}}(0)\rangle$
can be expressed as
\begin{eqnarray}
G_{\mathbf{\tilde{i}}\mathbf{\tilde{j}}}(\mathbf{k},i\omega)=\sum_{n}
\left(\frac{u^{\mathbf{k}}_{n,\mathbf{\tilde{i}},\uparrow}
u^{\mathbf{k}\ast}_{n,\mathbf{\tilde{j}},\uparrow}}{i\omega-E^{\mathbf{k}}_{n}}
+\frac{v^{\mathbf{k}}_{n,\mathbf{\tilde{i}},\downarrow}
v^{\mathbf{k}\ast}_{n,\mathbf{\tilde{j}},\downarrow}}{i\omega+E^{\mathbf{k}}_{n}}\right).
\end{eqnarray}
The spectral function $A(\mathbf{k},E)$ and the DOS $\rho(E)$ can be
derived respectively from the analytic continuation of the Green's
function as,
\begin{eqnarray}
A(\mathbf{k},E)=-\frac{1}{N_{P}\pi}\sum_{\mathbf{\tilde{i}}}\textmd{Im}
G_{\mathbf{\tilde{i}}\mathbf{\tilde{i}}}(\mathbf{k},iE\rightarrow
E+i0^{+}),
\end{eqnarray}
and
\begin{eqnarray}
\rho(E)=\frac{1}{N_{\mathbf{k}}}\sum_{\mathbf{k}}A(\mathbf{k},E),
\end{eqnarray}
where $N_{P}$ and $N_{\mathbf{k}}$ are the number of PUCs in the EUC
and the number of $\mathbf{k}$-points in the Brillouin zone,
respectively.

\section{results and discussion}

\subsection{Spectral weight distribution and energy gap
for the chiral flux phase} 
\begin{figure}[htb]
\begin{center}
\vspace{.2cm}
\includegraphics[width=230pt,height=190pt]{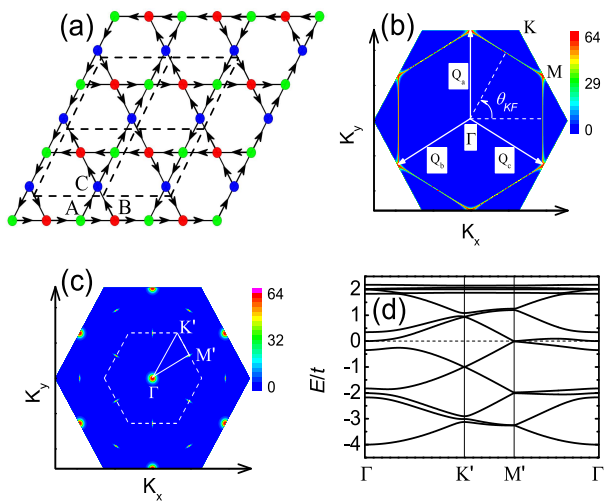}
\caption{(a) The lattice structure of the kagome superconductors,
made out of three sublattices $A$ (green dots), $B$ (red dots) and
$C$ (blue dots). The arrows on the lattice depict the configuration
of the orbital current, and the dashed lines denote the enlarged
unit cell. (b) Fermi surface produced by the spectral weight
distribution $A^{0}(\mathbf{k},E)$ at zero energy $E=0$. (c) The
spectral weight distribution $A(\mathbf{k},E)$ at zero energy $E=0$
obtained in the $2\times 2$ chiral flux phase with $\lambda=0.1$.
The hexagonal area in (c) surrounded by the white dashed lines shows
the reduced Brillouin zone in the $2\times 2$ chiral flux phase. (d)
The dispersion of the chiral flux phase along high-symmetry cuts in
the reduced Brillouin zone for a typical value of
$\lambda=0.1$.}\label{fig1}
\end{center}
\end{figure}

In this section, we investigate the
low-energy spectral weight distribution and the energy gap in the
chiral flux phase. In the calculations, the chemical potential $\mu$
is tuned so as to fix the band filling at $1/6$ hole doping, i.e.,
the van Hove filling, which facilitates the inter-scatterings
between three van Hove singularities with wave vectors
$\mathbf{Q}_{a,b,c}$. The scatterings of the charge order, on the
one hand, deplete the spectral weight on some portions of the FS, as
shown in Fig.~\ref{fig1}(c) for the low-energy ($E=0$) spectral
weight distribution that is calculated directly in the PBZ. On the
other hand, they fold the three bands in the PBZ of the PUC into the
reduced Brillouin zone (RBZ) of the EUC, forming the twelve bands in
the RBZ as presented in Fig.~\ref{fig1}(d). Nevertheless, the
results calculated straightly in the PBZ contain extra folded
segments of spectral weight and energy bands [see
Fig.~\ref{fig1}(c), Fig.~\ref{fig3}(b) and Fig.~\ref{fig6} for
reference], and thus hinder the direct comparison to the original
band structure or the experimental spectra in ARPES. Therefore, in
the following, we will employ an unfolding procedure to recover the
band structures and the spectral weights in the PBZ.

%%%%%%%%%%%%%%%%%%%%%%%%%%%%%%%%%%%%%%%
\vspace*{.2cm}
%%%%%%%%%%%%%%%%%%%%%%%%%%%%%%%%%%%%

\begin{figure}[htb]
\begin{center}
\vspace{.2cm}
\includegraphics[width=230pt,height=190pt]{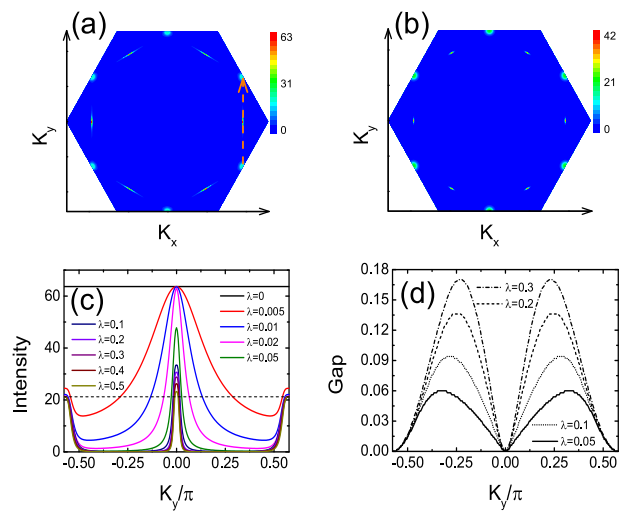}
\caption{The unfolded spectral weight distribution $A(\mathbf{k},E)$
at $E=0$ in the primitive Brillouin zone for $\lambda=0.02$ (a) and
$\lambda=0.1$ (b), respectively. The dashed orange arrow in (a)
indicates the momentum cut from $(\pi,-\sqrt{3}\pi/3)$ to
$(\pi,\sqrt{3}\pi/3)$ direction. (c) The momentum distribution
curves of the spectral weight along the momentum cut shown by the
dashed orange arrow in (a) for different $\lambda$. The dashed line
in (c) portrays one third of the intensity of $\lambda=0$. (d)
Evolution of the CDW gap along the momentum cut shown by the dashed
orange arrow in (a) for different $\lambda$. }\label{fig2}
\end{center}
\end{figure}

\begin{figure}[htb]
\begin{center}
\vspace{.2cm}
\includegraphics[width=230pt,height=196pt]{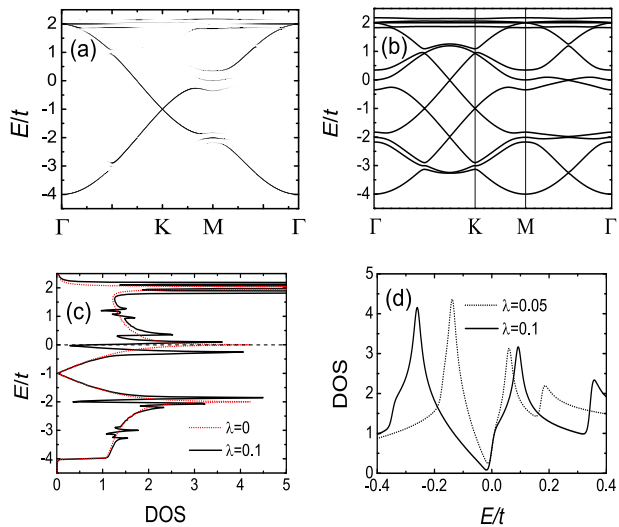}
\caption{(a) The unfolded dispersion of the chiral flux phase along
high-symmetry cuts in the primitive Brillouin zone for a typical
value of $\lambda=0.1$. (b) The folded energy band dispersion of the
chiral flux phase along high-symmetry cuts in the primitive
Brillouin zone for a typical value of $\lambda=0.1$. (c) Density of
states in a wide energy range for the normal state (the red dotted
curve) and the chiral flux state with $\lambda=0.1$ (the black solid
curve). The dashed line is the Fermi level corresponding to the van
Hove filling. (d) Density of states in the small energy scale for
$\lambda=0.05$ and $\lambda=0.1$ in the chiral flux
phase.}\label{fig3}
\end{center}
\end{figure}

In Figs.~\ref{fig2}(a) and (b), we show the unfolded spectral weight
distribution $A(\mathbf{k},E)$ at $E=0$ in the chiral flux phase for
$\lambda=0.02$ and $\lambda=0.1$, respectively. In order to get a
better view about the changes of the spectral weights for different
$\lambda$, Fig.~\ref{fig2}(c) also displays the unfolded spectral
distribution along the momentum cut from $(\pi,-\sqrt{3}\pi/3)$ to
$(\pi,\sqrt{3}\pi/3)$ direction. As can be seen from the figures,
the CDW scatterings not only depress the spectral weight and open
the CDW gap on the FS, but also make the depression exhibit distinct
momentum dependence. While the most depressed portion occurs near
the $M$ points for a weak $\lambda$ and extends toward the midpoint
of two adjacent $M$ points with the increase of $\lambda$, a finite
strength of spectral weight still remains around the midpoint and at
the $M$ points within the reasonable parameter regime $\lambda\leq
0.5$. Remarkably, the strength of spectral weight at the $M$ points
converges to a lower limit of one third of the intensity with the
increase of $\lambda$, and accordingly a round spot, much like a van
Hove singularity, with almost constant intensity of spectral weight
appears at the $M$ points.

%%%%%%%%%%%%%%%%%%%%%%%%%%%%%%%%%%%%%%%
\vspace*{.2cm}
%%%%%%%%%%%%%%%%%%%%%%%%%%%%%%%%%%%%

As a result of the momentum-dependent scatterings of the charge
order, the CDW gap acquires the strong momentum dependence.
Fig.~\ref{fig2}(d) presents the CDW gaps for different $\lambda$
along the momentum cut shown by the dashed orange arrow in
Fig.~\ref{fig2}(a), which are extracted from the spectral peak
positions in the momentum-energy space. It is obvious that the CDW
gap on the FS is anisotropic, with its zero minimum at the $M$
points and at the midpoint of two adjacent $M$ points, but
exhibiting a maximum in the middle of these two zero minimums. We
notice that the anisotropic CDW gap, including the positions of
minimum and maximum, agrees with the ARPES observations very
well~\cite{HLuo1}.

%%%%%%%%%%%%%%%%%%%%%%%%%%%%%%%%%%%%%%%
\vspace*{.2cm}
%%%%%%%%%%%%%%%%%%%%%%%%%%%%%%%%%%%%

Considering the fact that the three scattering wave vectors connect
the adjacent van Hove singular points ($M$ points), the zero-energy
van Hove singularity-like spot at the $M$ points in the CDW state
with the lower limit of intensity being exactly one third of that in
the normal state is very intriguing and worth further clarification.
It would be much more sensible to get insight from the physics of
the three van Hove singular points at van Hove filling, where each
of them comes exclusively from one of the three inequivalent lattice
sites in the normal state~\cite{SLYu1,Jiang1} and are mutually
coupled by the CDW order in the chiral flux phase. According to the
configuration of the orbital current in Fig.~\ref{fig1}(a), the
low-energy effective theory at the $M$ point for the CDW state is
well described by the patch model~\cite{Nand1,YPLin1,YPLin2},
\begin{eqnarray}
H_{CDW}(M)=\left(
\begin{array}{ccc}
\varepsilon_{M_{A}} & i\lambda &
i\lambda \\
-i\lambda & \varepsilon_{M_{B}} & -i\lambda \\
-i\lambda & i\lambda & \varepsilon_{M_{C}}
\end{array}
\right),
\end{eqnarray}
where $\varepsilon_{M_{A}}$ ($\varepsilon_{M_{B}}$,
$\varepsilon_{M_{C}}$) stands for the energy at the van Hove
singular point $M_{A}$ ($M_{B}$, $M_{C}$) that originates from the
sublattice $A$ ($B$, $C$). At the van Hove filling with
$\varepsilon_{M_{A}}=\varepsilon_{M_{B}}=\varepsilon_{M_{C}}=0$, one
can immediately find three eigenstates for the Hamiltonian
$H_{CDW}(M)$ with respective eigenvalues $E_{0}(M)=0$ and
$E_{\pm}(M)=\pm\sqrt{3}\lambda$, and each of these three eigenstates
weights one third of the total probability. This is to say the
zero-energy state at van Hove singular point $M$ triply splits into
three energy states with each of them possessing one third of the
total spectral weights.

Interestingly, the very simple argument is indeed embodied in the
momentum distribution curves of the unfolded spectral weight and the
unfolded energy dispersion in momentum space along high-symmetry
cuts shown respectively in Figs.~\ref{fig2}(c) and ~\ref{fig3}(a).
As illustrated in these figures, the original band near the upper
saddle point is triply split by the charge order scatterings to the
upper, middle and lower branches, where the middle one characterizes
a new saddle-shaped band just at the Fermi energy with its spectral
weight approaching one third of that in the normal state. By
contrast, the Dirac point around the $K$ point remains nearly intact
upon entering the chiral flux phase. For comparison, we also present
the folded energy band dispersion of the chiral flux phase along the
high-symmetry cuts in the PBZ in Fig.~\ref{fig3}(b). More detailed
energy band changes with $\lambda$ can be found in Appendix A. As a
result, the effect of a weak chiral flux charge order on the DOS
mainly concentrates on the near-by energies of the van Hove
singularity, as can be seen by making a comparison between the black
solid and the red dotted curves in Fig.~\ref{fig3}(c) and in
Fig.~\ref{fig3}(d) for the enlarged view of the low energy DOS. This
implies the orbital current order would have a profound impact on
the SC properties, which have been proposed to derive benefit from
the van Hove singularities.

Thus far, we have demonstrated that the proposed chiral flux CDW
phase for the kagome superconductors exhibits some typical features,
including the momentum-dependent partial gap opening at Fermi
energy, the emergence of the new saddle point at the $M$ point and
the unchanged Dirac point as well, which are all in good accordance
with the experimental
observations~\cite{PhysRevX.11.041010,HLuo1,SCho1}. It is worth
pointing out that the unique change of the spectral function from
the normal state to the CDW state can be used as an indirect
evidence to identify the time-reversal symmetry-breaking CDW state.
For the CDW state with time-reversal symmetry, the off-diagonal
elements in Eq. (12) are real, which leads to the band at $M$ point
splitting into only two sub-bands with unequal spectral weights.

%%%%%%%%%%%%%%%%%%%%%%%%%%%%%%%%%%%%%%%
\vspace*{.2cm}
%%%%%%%%%%%%%%%%%%%%%%%%%%%%%%%%%%%%
\begin{figure}[htb]
\begin{center}
\vspace{.0cm}
\includegraphics[width=210pt,height=230pt]{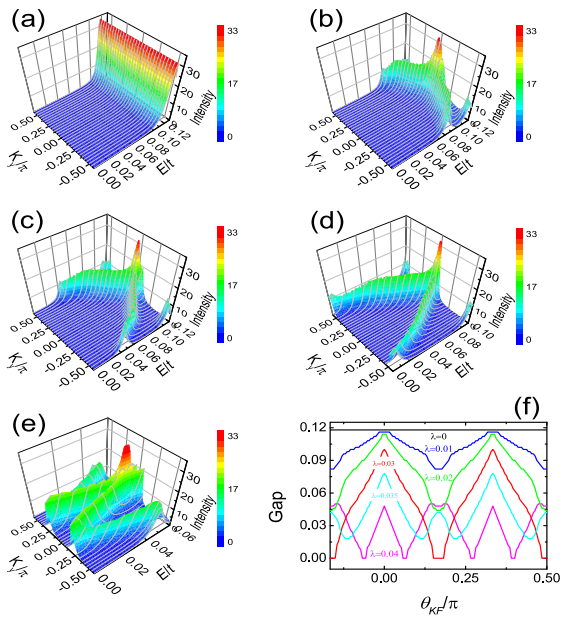}
\caption{The intensity plots of the unfolded spectral function as
functions of the momentum and energy in the coexistent state for
different strength of $\lambda=0$ (a), $\lambda=0.01$ (b),
$\lambda=0.02$ (c), $\lambda=0.03$ (d), and $\lambda=0.04$ (e),
respectively. The momentum in each panel is along the cut from
$(\pi,-\sqrt{3}\pi/3)$ to $(\pi,\sqrt{3}\pi/3)$ direction, as
denoted by the dashed orange arrow in Fig. 2(a). (f) Evolution of
the gap along the Fermi surface as a function of $\theta_{k_{F}}$
for different strength of $\lambda$. The gap is extracted from the
peak positions of the spectral function in panels
(a)-(e).}\label{fig4}
\end{center}
\end{figure}

\subsection{Modulated $s$-wave SC pairing and nodal gap}
\begin{figure}[htb]
\begin{center}
\vspace{.0cm}
\includegraphics[width=230pt,height=196pt]{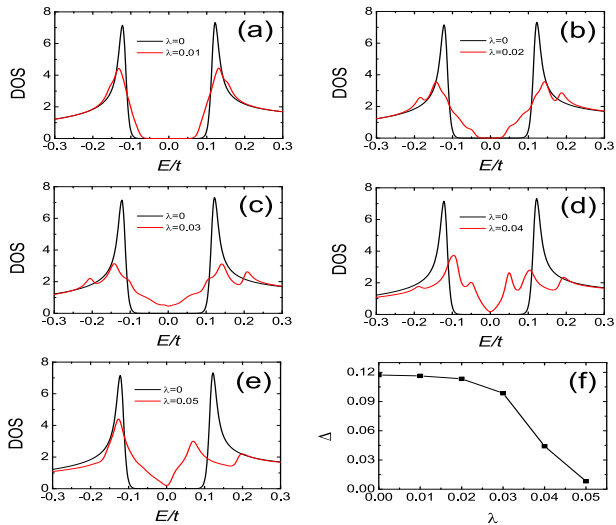}
\caption{(a)-(e) The energy dependence of the DOS for different
strength of $\lambda$. (f) Evolution of the SC pairing $\Delta$ as a
function of $\lambda$.}\label{fig5}
\end{center}
\end{figure}

Having analyzed the spectral and gap
features of the chiral flux CDW phase, we now pursue the main
question of the paper, i.e., the impact of the chiral flux order on
the SC properties. In the calculations, the SC order parameter is
determined self-consistently by treating the orbital current order
$\lambda$ as a variable argument. In Figs.~\ref{fig4}(a)-(e), we
present the intensity plots of the spectral function in the
coexistence of orbital current and SC orders as functions of energy
and momentum along the cut shown as the dashed orange arrow in
Fig.~\ref{fig2}(a). For $\lambda=0$ in Fig.~\ref{fig4}(a), the
intensity of the spectral function shows a gap edge structure at
constant energy, from which an isotropic gap feature indicated by
the black solid line in Fig.~\ref{fig4}(f) can be extracted on the
FS as a function of $\theta_{k_{F}}$, with the angle of the Fermi
momentum $\theta_{k_{F}}$ being defined in Fig.~\ref{fig1}(b).
Correspondingly, one can see a typical U-shaped full gap structure
for the DOS in Figs.~\ref{fig5}(a)-(e), where the DOS for
$\lambda=0$ is plotted with the black curve in all panels for
reference.

Inclusion of a small value of $\lambda$, say for example
$\lambda<0.02$, has little bearing on the SC pairing amplitude
$\Delta$ [see Fig.~\ref{fig5}(f)], but has an obvious effect on the
the spectral function distribution and the line shape of DOS.
Specifically, the introduction of a tiny value of the orbital
current order, such as $\lambda=0.01$, will cause the red-shift of
the gap edge. As the CDW order partially gaps the FS with the utmost
strength of the scattering occurring at the $M$ points, the SC
softening is expected to start from the $M$ points. This is
evidenced by the drawing in Fig.~\ref{fig4}(b), where the shift is
clearly visible to begin at the $M$ points, and reach its minimum at
the midpoint of the momentum cut, leading to the modulation of the
gap on the FS, as displayed by the blue curve in Fig.~\ref{fig4}(f).
As a result, the DOS changes its U-shaped curve to a basin-like one
along with the slightly blunted gap edges, as depicted by the red
curve in Fig.~\ref{fig5}(a).

Whereas the position of the gap edge at the midpoint of the momentum
cut shifts slightly toward lower energies with the increase of
$\lambda$, it decreases significantly at the $M$ points, as
illustrated in Figs.~\ref{fig4}(b)-(c). As is seen in
Fig.~\ref{fig4}(f), the striking contrast of the gap edge shifts
between the $M$ points and the midpoint results in the deeper
modulation depth of the spectral gap on the FS with a stronger
orbital current order. Consequently, one can expect the appearance
of the gap nodes on the FS when the orbital current order reaches a
certain level such as $\lambda\sim0.03$. Further increase in the
orbital current order will remove the gap nodes by lifting the
quasiparticle energies around the $M$ points on one hand, and on the
other hand it pushes the energies downward on the portion between
the $M$ point and the midpoint [see Fig.~\ref{fig4}(f) for
$\lambda=0.035$]. As a result, the period of the gap modulation on
the FS becomes shorter than that of the smaller $\lambda$, and leads
to the doubling of the nodal portions on the FS as the orbital
current order achieving to about $\lambda=0.04$, as is shown in
Fig.~\ref{fig4}(f). It is noteworthy that the results with sign
preserved nodal gap we identified here come to the same conclusion
with the experiment~\cite{HSXu1}.

With the increase of $\lambda$, the line shape of the DOS follows
the corresponding changes with the spectral weight function. On one
hand, accompanied by the decrease of the SC pairing amplitude
$\Delta$ [see Fig.~\ref{fig5}(f)], the basin-like DOS is continually
deformed to a V-shaped one by the increase of $\lambda$, as
evidenced in Figs.~\ref{fig5}(b)-(d). On the other hand, the
appearance of multiple sets of gap edge peak can be clearly seen in
Figs.~\ref{fig5}(b)-(d) for $\lambda\geq 0.02$. Moreover, as
$\lambda$ is increased, there exists residual DOS at zero energy in
Figs.~\ref{fig5}(c) and (d). The V-shaped DOS with multiple sets of
gap edge peak and residual zero-energy DOS constitutes a
characteristic of a nodal multi-gap SC pairing state. Although a
single orbital tight-binding model and the conventional on-site
$s$-wave SC pairing are adopted here, it is very interesting that
the salient features such as the V-shaped DOS, the residual
zero-energy DOS as well as the multiple sets of gap edge peak
produced in the coexistence of orbital current order and SC pairing
are in good accordance with the STM
experiments~\cite{HSXu1,Liang1,HChen1}.

%%%%%%%%%%%%%%%%%%%%%%%%%%%%%%%%%%%%%%%
\vspace*{.2cm}
%%%%%%%%%%%%%%%%%%%%%%%%%%%%%%%%%%%%%%%

As the orbital current order increases to $\lambda\sim 0.05$, the
system is driven to the phase of dominant CDW order by heavily
suppressing the SC pairing amplitude [see Fig.~\ref{fig5}(f)].
Accordingly, the DOS for $\lambda=0.05$ presented in
Fig.~\ref{fig5}(e) turns to the result in Fig.~\ref{fig3}(d) for the
pure chiral flux phase. The result suggests that the strength of the
orbital current order needs to be treated with caution when one
studies the interplay between the SC and orbital current order in a
non-self-consistent manner~\cite{YGu1}, because the
superconductivity would be totally suppressed by even the moderate
orbital current order.

Clarification of the interplay between the time-reversal
symmetry-breaking charge orders and superconductivity is a key step
toward the understanding of the underlying physics of the V-based
kagome superconductors. Although the competition nature between them
has been identified in these materials, there is accumulating
evidence that also shows a coexistence of them. It has been observed
that the CDW in AV$_{3}$Sb$_{5}$ intertwined with SC order could
result in the spatial modulations of the SC gap, i.e., the so called
roton pair-density wave state~\cite{HChen1}. In this paper, we have
demonstrated that the orbital current order can lead to the gap
modulations on the FS. While the spatial modulations of the gap have
been probed by means of STM/STS measurement, the gap modulations on
the FS can be discernable in the ARPES and field angle-dependent
thermal conductivity measurements for identifying the interplay
between the CDW and the SC orders, and for verifying or falsifying
the above scenario. Whether there is a relationship between the gap
modulations on the FS in the present study and the experimental
observations of the roton pair-density wave, or if there is a
possibility that the orbital current order could also lead to some
spatial modulations of the SC pairing, constitute the fascinating
questions deserving further studies.

\section {conclusion}
In conclusion, we have studied the impact of the chiral flux CDW
order on the experimental outcomes for the normal and the SC
properties. With the aids of the unfolding procedure to the PBZ, it
was revealed that a new saddle point turns up at $M$ points near the
Fermi energy to form the van Have singularity-like spectral spots in
the CDW phase, despite that the original band near the Fermi energy
could be gaped by the CDW order. The band gap was manifested in the
partially gaped spectral weight on the FS with the reservation of
the spectral weight on the $M$ points and the midpoint between the
two adjacent $M$ points. In the environment of the partially gaped
FS and the new van Have singularity-like spot, a conventional fully
gapped SC pairing would have a chance to survive in the coexistence
of the chiral flux CDW and SC orders, at the expense of the spectral
gap modulations on the FS. Owing to the modulations of the gap, a
nodal gap feature for the spectral weight might appear on the FS
under certain strengths of the CDW order. Accordingly, the U-shaped
DOS deformed to the V-shaped one along with the residual DOS near
the Fermi energy. These results were concordant with the
experimental observations in many aspects, and might serve as a
proposal to mediate the divergent or seemingly contradictory
experimental outcomes with respect to the SC pairing symmetry.

\section{acknowledgement}
\par
This work was supported by the National Natural Science Foundation
of China (Grant No. 12074175).

\appendix

\renewcommand{\thefigure}{\Alph{section}\arabic{figure}}

\begin{appendices}

\section{Folded and unfolded band structures for different $\lambda$}

In Fig.~\ref{fig6}, we present the folded band structures and the
unfolded dispersion of the spectral weight along high-symmetry cuts
in the PBZ for the chiral flux phase with different $\lambda$. First
of all, the triple splitting of the energy band near the van Hove
points happens for all cases with the splitting being proportional
to the charge order strength $\lambda$ (Note that the triple
splitting is obviously weakened at the lower van Hove point due to
the non-zero energy of this point and the mixing contributions from
different sublattices.). Secondly, the low energy portion along the
$\Gamma$ to $K$ direction remains intact when the CDW order is not
too strong such as $\lambda\leq 0.1$, but it is gapped by the strong
CDW order with $\lambda=0.2$ and $\lambda=0.3$ with the upper branch
just touching the Fermi level, which clearly manifest in
Figs.~\ref{fig6}(b), (d), (f) and (h) for the unfolded spectral
dispersion. Besides, the Dirac bands near the $K$ point always
maintain the same basic feature for the parameters considered here.

\setcounter{figure}{0}
\begin{figure}[h]
%\begin{adjustwidth}{-2.25in}{0in}
\begin{center}
{\mbox{\includegraphics[width=220pt,height=80pt]{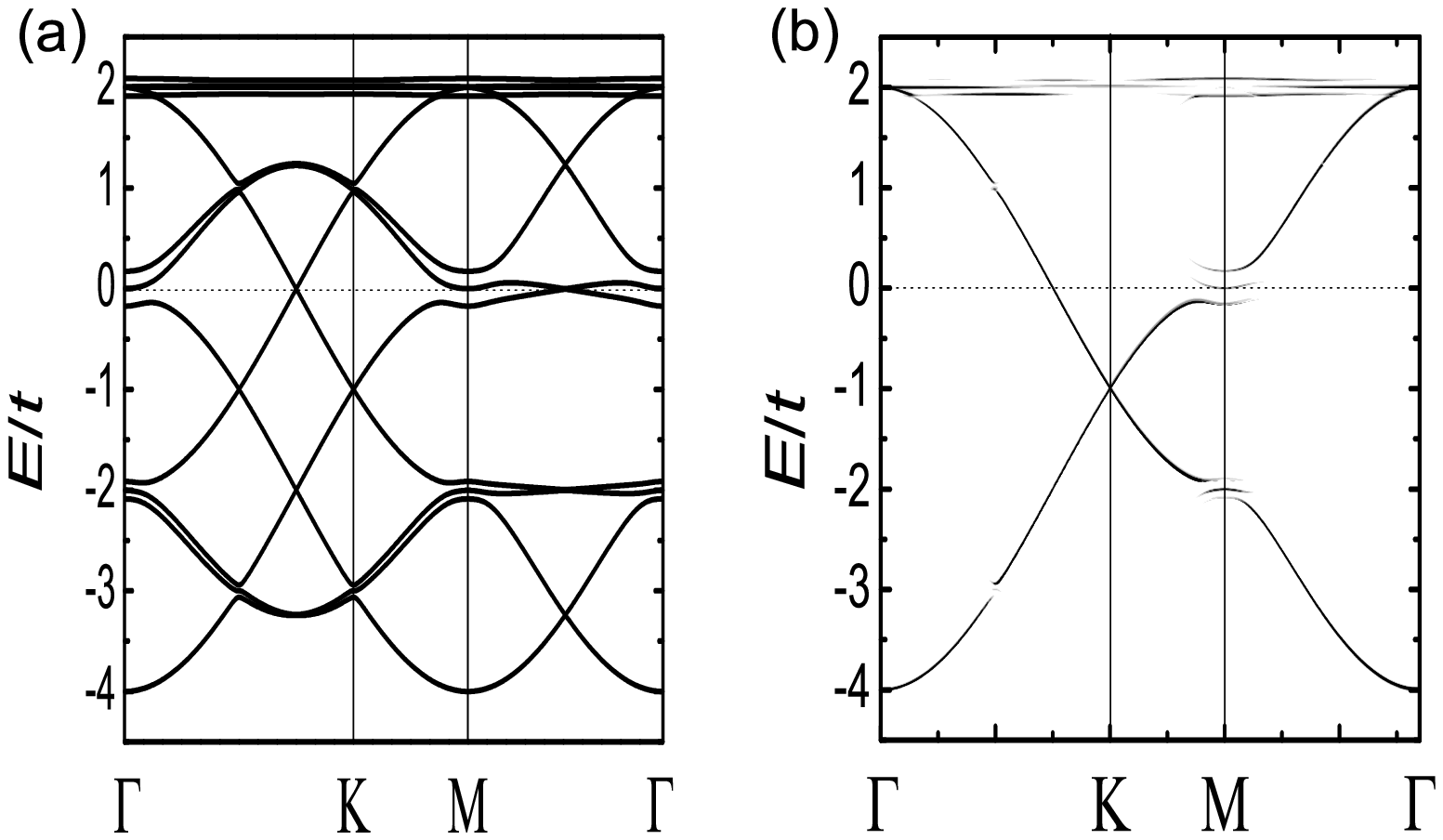}}}
\vspace{0.5cm}\\
{\mbox{\includegraphics[width=220pt,height=80pt]{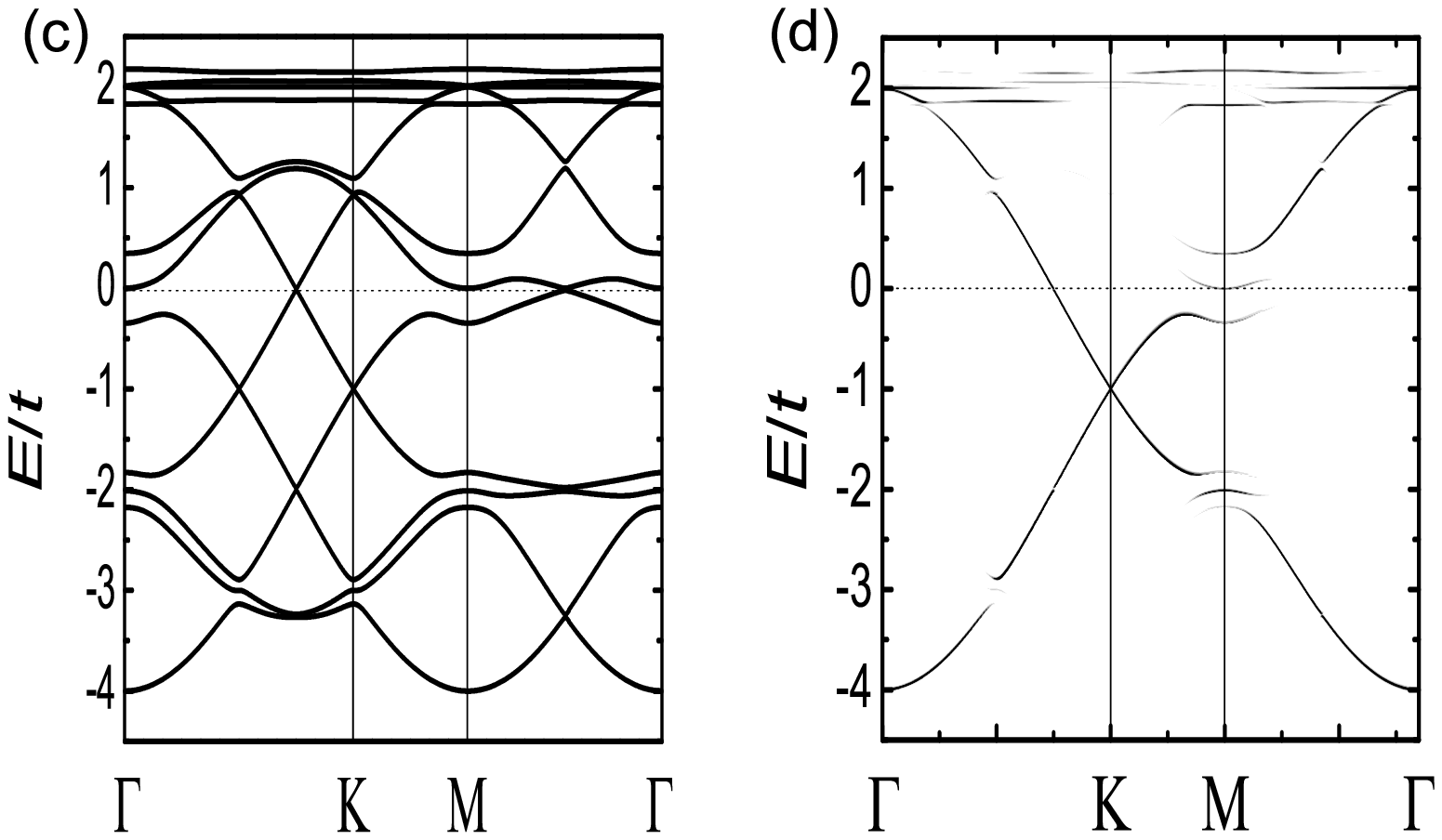}}}
\vspace{0.5cm}\\
{\mbox{\includegraphics[width=220pt,height=80pt]{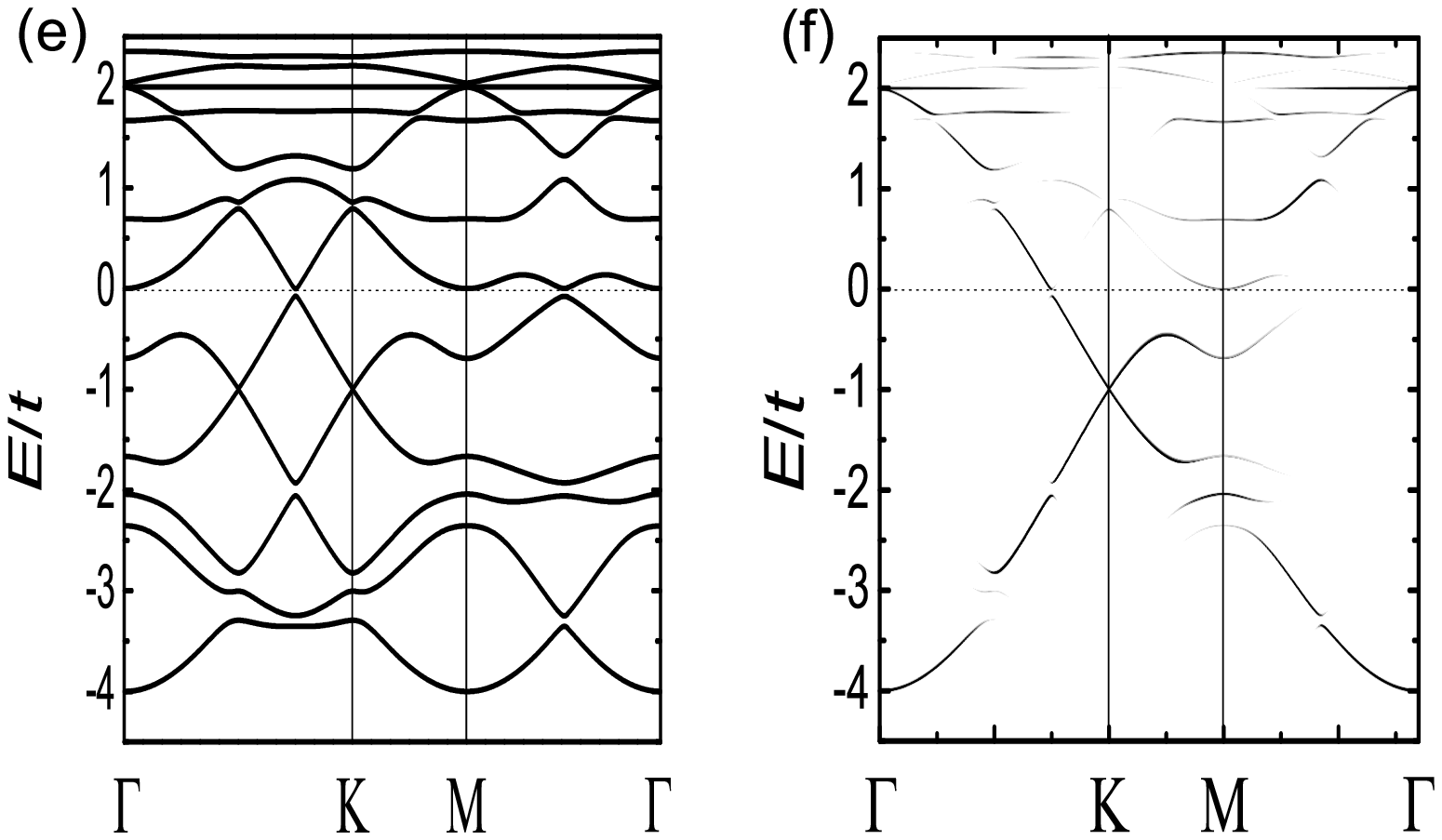}}}
\vspace{0.5cm}\\
{\mbox{\includegraphics[width=220pt,height=80pt]{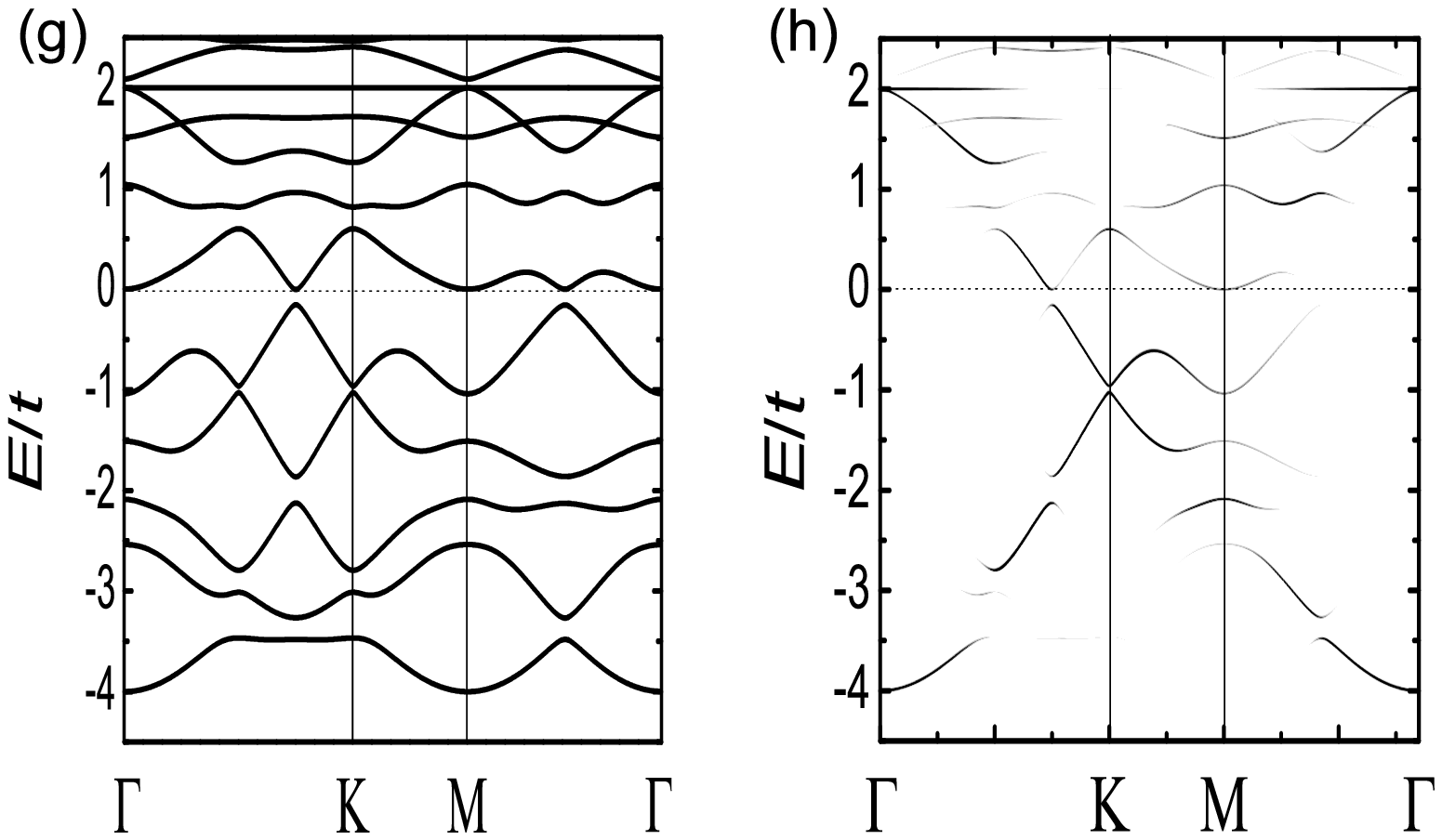}}}
\caption{The folded (left column) and unfolded (right column) band
structures of the chiral flux charge order phase along high-symmetry
cuts in the primitive Brillouin zone for $\lambda=0.05$ [(a), (b)],
$\lambda=0.1$ [(c), (d)], $\lambda=0.2$ [(e), (f)] and $\lambda=0.3$
[(g), (h)] respectively.} \label{fig6}
\end{center}
%\end{adjustwidth}
\end{figure}

\end{appendices}

\end{document}